\documentclass[
preprint,
eqsecnum,
amsmath,amssymb,
amsmath,amssymb, aps,
]{revtex4}
\usepackage[dvipsnames]{xcolor} 
\usepackage{epsfig}
\usepackage{amsfonts}
\usepackage{amsmath}
\usepackage{wasysym}
\usepackage{amssymb}
\usepackage{bm} 
\usepackage[framemethod=PStricks]{mdframed}
\usepackage{lipsum}
\usepackage{float}

\usepackage{pdflscape}
\usepackage{graphicx}
\usepackage{ulem}

\newcommand{\be}{\begin{equation}}
\newcommand{\ee}{\end{equation}} 
\newcommand{\bea}{\begin{eqnarray}} 
\newcommand{\eea}{\end{eqnarray}}

\usepackage[utf8]{inputenc}
\usepackage{bbm} 				
\usepackage{epstopdf}				
\usepackage{verbatim}    			
\usepackage{sidecap}                
\usepackage{indentfirst}
\usepackage[colorinlistoftodos]{todonotes}

\begin{document}







\maketitle



\centerline{\large{\bf{Minimal Surfaces Unveiled from the}}}
\centerline{\large{\bf{Statistics of Turbulent Circulation Fluctuations}}}
\vspace{0.5cm}

\centerline{Luca Moriconi}
\vspace{0.1cm}
\centerline{\it{Instituto de F\'\i sica, Universidade Federal do Rio de Janeiro,}} 

\centerline{\it{C.P. 68528, CEP: 21945-970, Rio de Janeiro, RJ, Brazil}}
\vspace{1.0cm}

Circulation is a key unifying concept in fluid dynamics \cite{acheson}. It plays a central role in a huge variety of phenomena, from hurricane dynamics 
and the aesthetically appealing vapor volutes that rise from a teacup to
the subtle physics of flight, not to mention a myriad of other fascinating instances \cite{lugt}. Simply formulated as
a linear functional of the velocity field $v_i = v_i ({\bf{x}},t)$,
\be
\Gamma = \oint_C v_i dx_i \ , \ \nonumber
\ee
circulation probes the swirling motions of the flow around a closed oriented contour $C$. In the case of turbulent regimes, however,
it has been notoriously hard to gather, from experiments or numerical simulations, detailed information on the statistical properties of circulation. Progress has been crucially tied to the solution of computational bottlenecks, related to issues of data storage capacity and processing speed. After years of hampered advances, a groundbreaking study is reported in PNAS. A massive numerical campaign carried out by Iyer et al. \cite{Iyer_etal_PNAS} brings to the spotlight a surprising connection between minimal surfaces \cite{cold_minicozzi,perez} and the statistical description of turbulent circulation fluctuations.

\begin{figure}[ht]
\hspace{0.0cm} \includegraphics[width=0.9\textwidth]{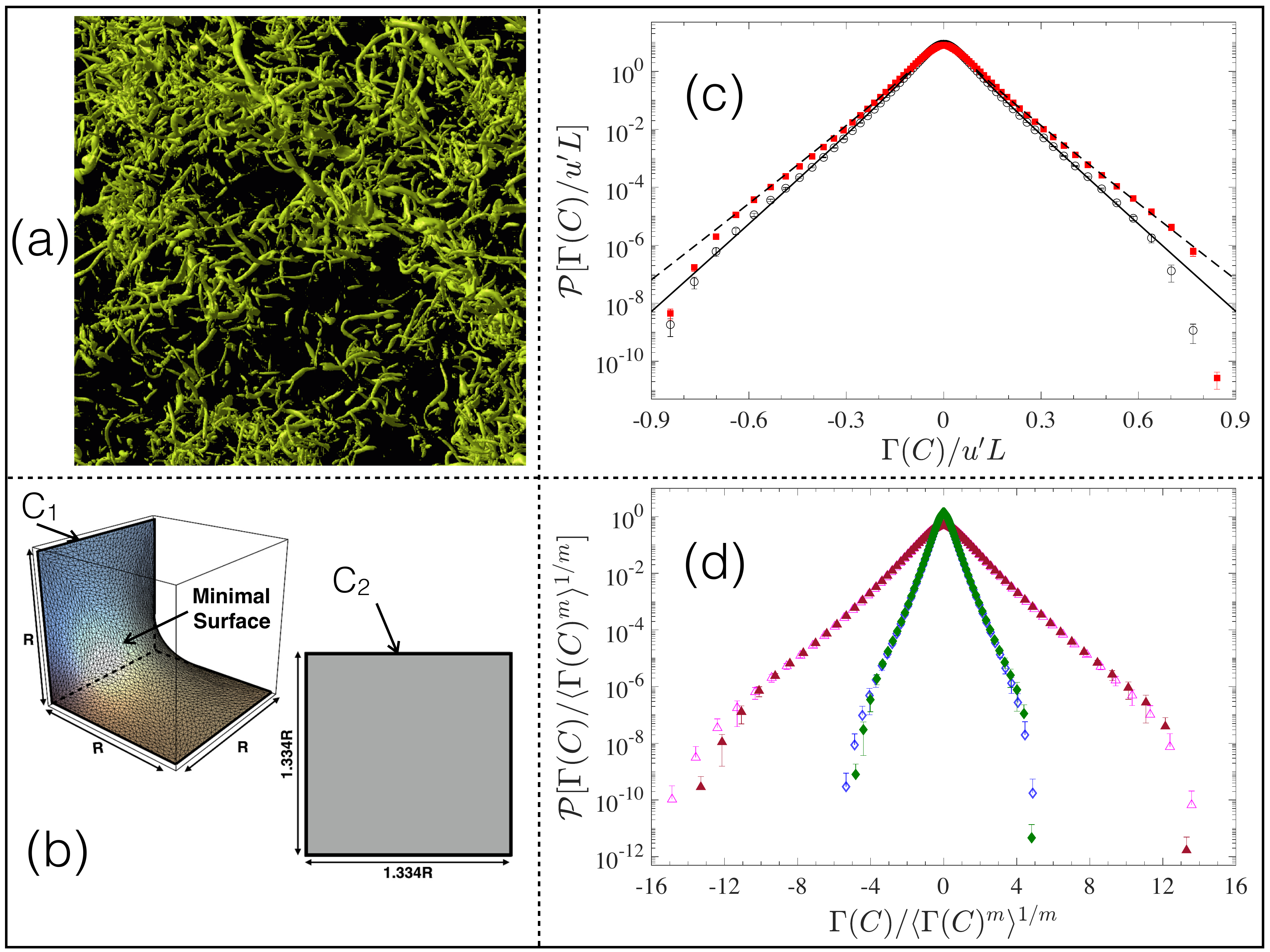}
\vspace{0.0cm}
\caption{(a) Worm-like vortex tubes as observed in the direct numerical simulations of homogeneous and isotropic turbulence by Yokokawa et al. \cite{yokokawa_etal}.
Reproduced with permission from  Ref. \cite{yokokawa_etal} (Copyright 2002, Institute of Electrical and Electronics Engineers); (b) $C_1$ (a non-planar loop) and $C_2$ (a planar loop) span the same minimal area and were used by Iyer et al. \cite{Iyer_etal_PNAS} to compute the cPDFs shown in (c) and (d). Open and filled symbols refer to the loops $C_1$ and $C_2$, respectively; (c) shows cPDFs which are rescaled by the same flow related, loop-independent, circulation scale; (d) circulation statistical moments of orders $m = 2$ (triangles) and $m = 8$ (diamonds) are taken to define loop-dependent circulation scales that lead to collapsing rescaled cPDFs. Figs. 1b, 1c, and 1d are adapted from Ref. \cite{Iyer_etal_PNAS}.} 
\label{}
\end{figure}

Since the mid-1990s, evidence has been accumulated to support the general picture of turbulent flows as systems of strongly coupled 
vortex structures \cite{she_etal,farge_etal,yokokawa_etal,yeung_etal} (Fig. 1a). In consonance with such phenomenological ideas, Migdal \cite{migdal1994,migdal2020} developed a statistical formalism for turbulence which has circulation as its main dynamical variable.
Relying upon formal similarities between the probability characteristic function of circulation and the Wilson loop observable of quantum chromodynamics (QCD), a 
gauge-invariant order parameter for the characterization of quark confined/deconfined phases \cite{wilson1974}, Migdal \cite{migdal1983} adapted to the domain of fluid dynamics the high-energy theoretical tool known as loop calculus. It is noteworthy to realize, in this connection, that one of the most promising attempts to model quark confinement stands on the existence of center vortices \cite{thooft1978,biddle_etal}, topological excitations of the gluon field that can be regarded as the QCD analogues of turbulent vortex tubes.  

The bottom line of the loop calculus’ approach to turbulence turned out to be the so-called area rule, a bold conjecture that asserts that the tails of circulation probability distribution functions (cPDFs) should depend, at asymptotically high Reynolds numbers, uniquely on the area spanned by the minimal surface bounded by the loop contour \cite{migdal1994,migdal2020}. 

The extensive numerical analysis of the area rule conjecture provided by Iyer et al. \cite{Iyer_etal_PNAS} is a computational tour de force, with its starting point dating back to 2019 \cite{Iyer_etal2019}. In that preliminary study, intriguing scaling properties of turbulence where discovered, as a transitional behavior for the statistical moments of circulation as their orders are varied.
That work triggered activity in a research horizon that encompasses a possible vortex structural approach to the multifractal nature of the turbulent cascade \cite{apol_etal,mori} and the close phenomenological connections between classical and quantum turbulence \cite{skr_sreeni, muller_etal}. In addition, initial prospects seemed to favor the correctness of the area rule conjecture, but more detailed and comprehensive analyses were in order. 

As the product of subsequent careful explorations \cite{Iyer_etal_PNAS}, the area rule is seen not to hold in a strict sense for general (planar or non-planar) loops (Fig. 1b and c), even though it could be sustained as a reasonable approximation for the case of planar loops. In view of these findings, the downfall of the area rule seemed to be inevitable, resonating with the natural guess that cPDF tails could have their shapes (i.e., functional forms) arbitrarily determined by specific loop details. Notwithstanding such worrying perspectives, here comes the plot twist. Further striking observations indicated that minimal surfaces do have a special place in turbulent circulation phenomenology. Actually, it was found that
\vspace{0.2cm}

\noindent $\bullet$ the decay of the cPDF tails for an arbitrary loop $C$ is well modeled by a simple exponential function (modulated by a $1/\sqrt{\Gamma}$ prefactor) characterized by a single circulation scale $\bar \Gamma_C$ (facts that agree well with predictions put 
forward in Ref. \cite{migdal2020})
\vspace{0.2cm}

\noindent and that 
\vspace{0.2cm}

\noindent $\bullet$ if the circulation variable is expressed in units of $\bar \Gamma_C$, the rescaled (dimensionless) cPDF tails collapse to a standardized shape which is loop-independent for a given minimal surface area, as it can be remarkably noticed from Fig. 1d.
\vspace{0.2cm}

The modified area rule empirically established in Ref. \cite{Iyer_etal_PNAS} as the second of the above two items is elegantly simple and truly unexpected. The rescaled cPDFs could depend on the specific loop details in an infinity of ways, but, against all odds, they do not. In synthesis, the shape of properly rescaled cPDF tails is expected to depend, at a fixed Reynolds number and within a broad range of scales, only on $A/L^2$, the ratio of the minimal surface area $A$ spanned by the loop to the square of the integral length scale of the turbulent flow, $L$.

There is a good deal of mathematical wonder encoded in the fancy metastable shapes of soap bubble wires, which can be regarded as true minimal surface factories \cite{perez}. Phenomenological bridges to minimal surface theory have been raised from the physics of QCD \cite{thooft1978, biddle_etal}, several branches of chemistry \cite{andersson_etal}, and porous media \cite{lin_etal}. This list of applications is concretely enlarged, from now on, with turbulence. 

The results of Iyer et al. \cite{Iyer_etal_PNAS} open important gates for the development of innovative research along theoretical, numerical, and experimental roads. Circulation is the right observable to explore the dynamics of a complex system of interacting vortex structures, which by means of whimsical tricks still to be clarified, provokes the curious mind with the mathematical beauty of minimal surfaces.


\begin{references}

\bibitem{acheson} D.J. Acheson, Elementray Fluid Dynamics, Oxford University Press (1998).

\bibitem{lugt} H.J. Lugt, Vortex Flows in Nature and Technology, Krieger Publishing Company, 
Malabar, Florida (1995).

\bibitem{Iyer_etal_PNAS} K.P. Iyer, S.S. Bharadwaj, and K.R. Sreenivasan, The area rule for circulation in three-dimensional turbulence, Proc. Nat. Acad. Sci. U.S.A. {\bf{118}}, e2114679118 (2021).

\bibitem{cold_minicozzi} T.H. Colding and W.P. Minicozzi II, A Course in Minimal Surfaces, Graduate Texts in Mathematics 121, The Americal Mathematical Society (2011).

\bibitem{perez} J. Pérez, A New Golden Age of Minimal Surfaces, Not. Am. Math. Soc. {\bf{64}}, 347 (2017).

\bibitem{she_etal} Z.-S. She, E. Jackson, and S.A. Orszag, Intermittent vortex structures in homogeneous
isotropic turbulence, Nat. {\bf{344}}, 226 (1990). 

\bibitem{farge_etal} M. Farge, G. Pellegrino, and K. Schneider, Coherent Vortex Extraction in 3D Turbulent Flows Using Orthogonal Wavelets,Phys. Rev. Lett. {\bf{87}}, 054501 (2001).

\bibitem{yokokawa_etal} M. Yokokawa, K. Itakura, A. Uno, T. Ishihara, and Y. Kaneda, in Supercomputing, ACM/IEEE 2002 Conference (2002).

\bibitem{yeung_etal} P.K. Yeung, X.M. Zhai, and K.R. Sreenivasan, Extreme events in computational turbulence, Proc. Nat. Acad. Sci. {\bf{112}}, 12633 (2015).

\bibitem{migdal1994} A.A. Migdal, Loop equation and area law in turbulence, Int. J. Mod. Phys. A {\bf{9}}, 1197 (1994).

\bibitem{migdal2020} A. Migdal, Clebsch confinement and instantons in turbulence, Int. J. Mod. Phys. A {\bf{35}}, 2030018 (2020).

\bibitem{wilson1974} K.G. Wilson, Confinement of Quarks, Phys. Rev. D {\bf{10}}, 2445 (1974).

\bibitem{migdal1983} A.A. Migdal, Loop Equations and $1/N$ Expansion, Phys. Rep. {\bf{102}}, 199 (1983).

\bibitem{thooft1978} G.’t Hooft, On the Phase Transition Towards Permanent Quark Confinement, Nucl. Phys. B {\bf{138}}, 1 (1978).

\bibitem{biddle_etal} C. Biddle, W. Kamleh, and D.B. Leinweber, Visualization of center vortex structure, Phys. Rev. D {\bf{102}}, 034504 (2020).

\bibitem{Iyer_etal2019} K.P. Iyer, K.R. Sreenivasan, and P.K. Yeung, Circulation in High Reynolds Number Isotropic Turbulence is a Bifractal, Phys. Rev. X {\bf{9}}, 041006 (2019).


\bibitem{apol_etal} G.B. Apolinário, L. Moriconi, R.M. Pereira, and V.J. Valadão, Vortex gas modeling of turbulent circulation statistics, Phys. Rev. E {\bf{102}}, 041102(R) (2020).

\bibitem{mori} L. Moriconi, Multifractality breaking from bounded random measures, Phys. Rev. E {\bf{103}}, 062137 (2021).

\bibitem{skr_sreeni} L. Skrbek and K.R. Sreenivasan, How Similar is Quantum Turbulence to Classical Turbulence?, in Ten Chapters in Turbulence, Cambridge University Press (2012).

\bibitem{muller_etal}  N.P. Müller, J.I. Polanco, and G. Krstulovic, Intermittency of Velocity Circulation in Quantum Turbulence, Phys. Rev. X {\bf{11}}, 011053 (2021).

\bibitem{andersson_etal}  S. Andersson, S.T. Hyde, K. Larsson, and S. Lidin, Minimal surfaces and structures: from inorganic and metal crystals to cell membranes and biopolymers, Chem. Rev. {\bf{88}}, 221 (1988).

\bibitem{lin_etal} Q. Lin, B. Bijeljic, S. Berg, R. Pini, M.J. Blunt, and S. Krevor, Minimal surfaces in porous media: Pore-scale imaging of multiphase flow in an altered-wettability Bentheimer sandstone, Phys. Rev. E {\bf{99}}, 063105 (2019).


\end{references}
\end{document}